\documentstyle[aps,tighten]{revtex}

\newcommand{\Christoffel}[2]{{\textstyle\left\{\! {#1 \atop #2}\! \right\}} }

\begin{document}

\draft

\preprint{UTPT-94-38}

\title{Test-Particle Motion in the
Nonsymmetric Gravitational Theory}

\author{J.\ L\'egar\'e and J.\ W.\ Moffat}

\address{Department of Physics, University of Toronto,
Toronto, Ontario, Canada M5S 1A7}

\date{May 30, 1995}

\maketitle

\begin{abstract}

We present a derivation of the equation of motion for a
test-particle in the framework of the nonsymmetric
gravitational theory.
Three possible couplings of the test-particle to the
non-symmetric gravitational field are explored.
The equation of motion is found to be similar in form
to the standard geodesic equation of general relativity,
but with an extra antisymmetric force term present.
The equation of motion is studied for the case of a
static, spherically symmetric source, where the extra force
term is found to take the form of a Yukawa force.

\end{abstract}


\section{Introduction}

We consider here the problem of test-particle motion in the
weak-field limit of the
nonsymmetric gravitational theory (NGT)
(see \cite{bib:Moffat_NGT_2} for
a description of the structure of the NGT).
The problem is approached from the point of view of Lagrangian
mechanics: a scalar Lagrangian that couples particle quantities
to the components of the gravitational field is postulated, and the
Euler-Lagrange equation is used to determine the equation
of motion.

In general relativity (GR), the scalar Lagrangian of choice describing
the coupling of particle quantities to the gravitational field is
well-known: $L_{\rm GR} = (g_{\mu\nu} \dot\xi^\mu\dot\xi^\nu)^{1/2}$.
However, the metric in GR is a symmetric tensor
$g_{\mu\nu} = g_{(\mu\nu)}$.
In the NGT, this is no longer the case, as the metric picks up a
skew component.
A problem therefore arises: if we describe the coupling of the
test-particle to the NGT gravitational field simply by $L_{\rm GR}$,
particle quantities will not couple to the skew components of the metric.
This leads to a more serious problem: if particles do not couple to
the skew components of the metric, how did the skew components come
to be?

The program therefore
consists in finding a Lagrangian which describes the
coupling of the particle quantities to both the symmetric (GR)
components of the gravitational field, as well as to the nonsymmetric
(NGT) elements of this field.
We present three possible couplings herein.
In formulating these Lagrangians, we have restricted ourselves to
couplings that are linear in the velocities.

\section{Test-Particle Motion}

As in GR, spacetime in the NGT is described by a four dimensional manifold
$M$ and a metric $g_{(\mu\nu)}$.
However unlike GR, the NGT field equations not only determine the
symmetric components $g_{(\mu\nu)}=s_{\mu\nu}$ of the metric, but also the
antisymmetric components $g_{[\mu\nu]}=a_{\mu\nu}$.
Let $s^{\mu\nu}$ be defined
by $s^{\mu\nu} s_{\mu\beta} = \delta^\nu_\beta$.
Note that $s^{\mu\nu} \ne g^{(\mu\nu)}$, and similarly,
$a^{\mu\nu} \ne g^{[\mu\nu]}$.
The inverse of the full metric is defined by $g^{\mu\nu}g_{\alpha\nu}
=g^{\nu\mu}g_{\nu\alpha}=\delta^\mu_\alpha$.

The equation of particle motion is to be derived from a scalar
Lagrangian $L$.
This Lagrangian will be decomposed into two pieces: $L = L_{\rm GR}
+ L_{\rm NGT}$,
where
\begin{equation}
\label{eq:kinetic_Lagrangian}
L_{\rm GR} = ( s_{\mu\nu}
\dot\xi^\mu \dot\xi^\nu)^{1/2} .
\end{equation}
Here, $\sigma$ is an affine parameter,
while $\xi^\mu(\sigma)$ and $\dot\xi^\mu(\sigma)=d\xi^\mu/d\sigma$
are the particle's path
and four-velocity (with respect to the affine parameter $\sigma$),
respectively.
We will take $L_{\rm NGT}$ to have the form
$L_{\rm NGT}= (1/2)\lambda A_\mu \dot\xi^\mu$,
where $A_\mu$ is a covector independent of the
particle velocity $\dot\xi^\mu$, and $\lambda$ is a coupling constant.
The Euler-Lagrange equation of particle mechanics
\begin{equation}
\label{eq:EL-equation}
\frac{d}{d\sigma}\frac{\partial L}{\partial\dot\xi^\alpha}
- \frac{\partial L}{\partial\xi^\alpha}
= 0
\end{equation}
allows us to conclude that the contribution of $L_{\rm NGT}$ to the
equation of motion will be of the form
\begin{equation}
\label{eq:skew_contribution}
\frac{d}{d\sigma}\frac{\partial L_{\rm NGT}}{\partial\dot\xi^\alpha}
- \frac{\partial L_{\rm NGT}}{\partial\xi^\alpha}
= \frac{\lambda}{2}\left(\frac{dA_\alpha}{d\sigma}
- \dot\xi^\mu\frac{\partial A_\mu}{\partial\xi^\alpha}\right)
= \lambda \dot\xi^\mu \partial_{[\mu} A_{\alpha]}
= - \lambda f_{[\alpha\mu]} \dot\xi^\mu ,
\end{equation}
where $f_{[\alpha\mu]} = \partial_{[\alpha}A_{\mu]}$, and where
we have used the fact that $d/d\sigma =
\dot\xi^\mu\partial/\partial\xi^\mu$.

Let $L_{\rm NGT} = L_1+L_2+L_3$, where
\begin{mathletters}
\label{eq:NGT_Lagrangians}
\begin{eqnarray}
L_1 &=&
\label{eq:dual_coupling}
\frac{1}{2} C_1 \lambda {}^*\!F^\eta\dot\xi_\eta =
\frac{C_1 \lambda}{2}
\frac{\epsilon^{\mu\nu\lambda\eta}}{\sqrt{-g}}
F_{[\mu\nu\lambda]}s_{\eta\sigma}\dot\xi^\sigma \\
L_2 &=&
\label{eq:direct_coupling}
\frac{1}{2} C_2 \lambda g^{[\mu\nu]}
F_{[\mu\nu\lambda]} \dot\xi^\lambda \\
L_3 &=&
\label{eq:skew_divergence}
\frac{C_3 \lambda}{2}
\frac{{{\bf g}^{[\lambda\nu]}}_{,\nu}}{\sqrt{-g}}
s_{\lambda\mu}\dot\xi^\mu .
\end{eqnarray}
\end{mathletters}The
field-strength tensor $F_{[\mu\nu\lambda]}$ is defined by
\begin{equation}
\label{eq:curl_skew_g}
F_{[\mu\nu\lambda]}
= \partial_{[\lambda}g_{\mu\nu]}
= \frac{1}{3}(\partial_\lambda a_{\mu\nu}
+ \partial_\mu a_{\nu\lambda}
+ \partial_\nu a_{\lambda\mu}) .
\end{equation}
The constant $\lambda$ couples the test-particle to the NGT
skew field, and has dimensions of a length.
The constants $C_i$ ($i=1,2,3$) are dimensionless constants
measuring the relative strengths of the three interactions.
The symbol $\epsilon^{\mu\nu\lambda\eta}$ is the fully antisymmetric
Levi-Civita tensor density, defined by
\[
\epsilon^{\mu\nu\lambda\eta} = \left\{
\begin{array}{cl}
+1 & \mbox{if $\mu\nu\lambda\eta$ is an even permutation of $1230$,} \\
-1 & \mbox{if $\mu\nu\lambda\eta$ is an odd permutation of $1230$,} \\
0 & \mbox{otherwise.}
\end{array}
\right.
\]
We use the notation ${\bf T}_{\mu\nu} = \sqrt{-g}T_{\mu\nu}$, for some
tensor $T_{\mu\nu}$.
All indices are lowered with $s_{\mu\nu}$ and raised with
$s^{\mu\nu}$.

We are now in a position to use the covariant Euler-Lagrange equation
of particle mechanics
to derive the equation of motion.
Inserting (\ref{eq:kinetic_Lagrangian}) and
(\ref{eq:NGT_Lagrangians}) into (\ref{eq:EL-equation})
and using (\ref{eq:skew_contribution}) gives
\begin{equation}
\label{eq:first_step}
0 = \frac{d\dot\xi^\beta}{d\sigma}
+ s^{\beta\alpha}\left(\dot\xi^\nu \frac{ds_{\alpha\nu}}{d\sigma}
- \frac{1}{2}s_{\mu\nu,\alpha}\dot\xi^\mu\dot\xi^\nu\right)
- \kappa\lambda s^{\beta\alpha} f_{[\alpha\mu]}
\dot\xi^\mu ,
\end{equation}
where we have contracted with $s^{\beta\alpha}$ and
where $\kappa^2 = s_{\mu\nu}\dot\xi^\mu \dot\xi^\nu$.
We have assumed that $\kappa$ is a
constant;
this assumption will be justified in the next section.
$f_{[\alpha\mu]}$ is given by
\begin{equation}
\label{eq:skew_force}
f_{[\alpha\mu]}
= C_1 \partial_{[\alpha}
\left(\frac{\epsilon^{[\sigma\nu\lambda\eta]}}{\sqrt{-g}}
F_{[\sigma\nu\lambda]} s_{\mu]\eta}\right)
+ C_2 \partial_{[\alpha}
\left(g^{[\eta\nu]}F_{[\eta\nu\mu]]}\right)
+ C_3 \partial_{[\alpha}\left(\frac{{{\bf g}^{[\eta\nu]}}_{,\nu}}
{\sqrt{-g}} s_{\mu]\eta}\right) .
\end{equation}
If we define the symmetric Christoffel symbols by
\begin{equation}
\label{eq:Christoffel_symbol}
\Christoffel{\beta}{(\mu\nu)}
= \frac{1}{2}s^{\beta\alpha}(s_{\alpha\nu,\mu}
+ s_{\mu\alpha,\nu} - s_{\mu\nu,\alpha}) ,
\end{equation}
we can rewrite (\ref{eq:first_step}) as
\begin{equation}
\label{eq:equation_of_motion}
\frac{d^2\xi^\beta}{d\sigma^2}
+ \Christoffel{\beta}{(\mu\nu)}
\frac{d\xi^\mu}{d\sigma}\frac{d\xi^\nu}{d\sigma}
= \kappa \lambda s^{\beta\alpha}f_{[\alpha\mu]}
\frac{d\xi^\mu}{d\sigma} .
\end{equation}

\section{Conservation Laws}
\label{sec:Conservation_Laws}

Contracting (\ref{eq:equation_of_motion}) with $s_{\alpha\beta}X^\alpha$
where $X^\alpha$ is some vector, gives
\[
\frac{d}{d\sigma}(s_{\alpha\beta}X^\alpha\dot\xi^\beta)
= \frac{d(X_\beta \dot\xi^\beta)}{d\sigma}
= \frac{1}{2}\dot\xi^\mu\dot\xi^\nu X^\alpha\partial_\alpha s_{\mu\nu}
+ s_{\alpha\mu}\dot\xi^\mu\dot\xi^\nu\partial_\nu X^\alpha
+ \kappa\lambda f_{[\alpha\mu]}\dot\xi^\mu X^\alpha .
\]
However, the Lie derivative of the symmetric part of the metric is
given by
\[
\pounds_X[s]_{\mu\nu} = X^\alpha\partial_\alpha s_{\mu\nu}
+ s_{\alpha\mu}\partial_\nu X^\alpha + s_{\alpha\nu}\partial_\mu X^\alpha.
\]
Therefore,
\begin{equation}
\label{eq:eq_of_motion_with_vector}
\frac{d(X_\beta\dot\xi^\beta)}{d\sigma}
= \frac{1}{2} \dot\xi^\mu \dot\xi^\nu \pounds_X[s]_{\mu\nu}
+ \kappa\lambda f_{[\alpha\mu]}\dot\xi^\mu X^\alpha .
\end{equation}

If in (\ref{eq:eq_of_motion_with_vector}) we take $X^\alpha$ to be the
components of a Killing vector, we have that
\[
\frac{d(X_\beta\dot\xi^\beta)}{d\sigma}
= \kappa\lambda f_{[\alpha\mu]}\dot\xi^\mu X^\alpha .
\]
If the right-hand side of this equation vanishes, we have the result
familiar from GR that $X_\beta \dot\xi^\beta$ is a constant of the
motion.
For instance, it will be shown below that the only non-vanishing
component of $f_{[\alpha\mu]}$ for a static, spherically-symmetric
system is $f_{[rt]}$.
Since $X = \partial/\partial\phi$ is a Killing vector of a
spherically-symmetric system, we find that
$s_{\phi\beta}\dot\xi^\beta = r^2\dot\phi\sin\theta$ is a constant of
the motion, corresponding to the conservation of angular momentum.

This is not a generic feature: in a static, spherically-symmetric
system, $X=\partial/\partial t$ is also a Killing vector, but since
$f_{[rt]}\ne 0$, then $s_{t\beta}\dot\xi^\beta = \gamma \dot t$ is not
a conserved quantity.
The nature of $\gamma\dot t$ in a static, spherically-symmetric system
will be explored further in the next section,
where it will be shown that $\gamma\dot t$
is one part of a conserved quantity.

Setting $X^\alpha=\dot\xi^\alpha$ in (\ref{eq:eq_of_motion_with_vector})
leads to
\[
\frac{d}{d\sigma}(s_{\alpha\beta}\dot\xi^\alpha\dot\xi^\beta)
= \frac{1}{2} \dot\xi^\mu \dot\xi^\nu \pounds_{\dot\xi}[s]_{\mu\nu}
+ \kappa\lambda f_{[\alpha\mu]}\dot\xi^\mu \dot\xi^\alpha .
\]
Now,
\[
\pounds_{\dot\xi}[s]_{\mu\nu}\dot\xi^\mu\dot\xi^\nu
= \dot\xi^\mu\dot\xi^\nu \frac{ds_{\mu\nu}}{d\sigma}
+ 2s_{\mu\nu} \dot\xi^\nu \frac{d\dot\xi^\mu}{d\sigma}
= 2\kappa\lambda f_{[\nu\mu]}\dot\xi^\nu\dot\xi^\mu \equiv 0 .
\]
Evidently, the last term in (\ref{eq:eq_of_motion_with_vector})
vanishes identically, leaving
\begin{equation}
\label{eq:conservation_of_mass}
\frac{d}{d\sigma} (s_{\alpha\beta}\dot\xi^\alpha\dot\xi^\beta)
= \frac{d\kappa^2}{d\sigma} = 0 .
\end{equation}
This justifies our assumption that $\kappa$ is a constant.
In fact, $\kappa$ is a first-integral of the motion, corresponding
physically to the constancy of the mass of the test-particle.

In GR, massive particles are usually assumed to have
$\kappa^2 = s_{\mu\nu} \dot\xi^\mu \dot\xi^\nu \ne 0$.
Indeed, it is generally assumed that $\kappa^2 = 1$, which defines
the affine parameter as the proper time: $d\tau^2
= \kappa^2 \, d\sigma^2$.
Then, the equation of motion becomes
\begin{equation}
\label{eq:equation_of_motion_massive}
\frac{d^2\xi^\beta}{d\tau^2}
+ \Christoffel{\beta}{(\mu\nu)}
\frac{d\xi^\mu}{d\tau}\frac{d\xi^\nu}{d\tau}
= \lambda s^{\beta\alpha}f_{[\alpha\mu]}
\frac{d\xi^\mu}{d\tau} .
\end{equation}

On the other hand, massless particles (such as
photons) are usually taken to have
$\kappa^2 = 0$.
Since the mass of the particle does not enter into
(\ref{eq:equation_of_motion}), we might be tempted to postulate that
this is the correct equation of motion for a massless particle.
However, for such a massless particle, the right hand side of
(\ref{eq:equation_of_motion}) vanishes, since $\kappa^2 = 0$.
The equation of motion for a massless particle in the NGT is
therefore written:
\begin{equation}
\label{eq:equation_of_motion_massless}
\frac{d^2\xi^\beta}{d\sigma^2}
+ \Christoffel{\beta}{(\mu\nu)}
\frac{d\xi^\mu}{d\sigma}\frac{d\xi^\nu}{d\sigma} = 0 .
\end{equation}
This is, of course, the geodesic equation of GR.
We therefore conclude that massless test-particles
in this model do not couple
directly to the antisymmetric components of the NGT gravitational
field, but rather follow geodesics of the symmetric background
$s_{\mu\nu}$.
This is equivalent to the statement that massless particles have no
structure.

\section{Massive Test-Particle Motion in a
Static Spherically Symmetric Field}

We now consider the motion of a test-particle in the gravitational field
of a static, spherically symmetric source.
We will restrict ourselves to the case of massive test-particles.
The coordinates are $\xi^1=r$, $\xi^2=\theta$,
$\xi^3=\phi$ and $\xi^0=t$.
For the case of ``non-magnetic'' NGT considered here, the
spherically-symmetric metric takes the form
\begin{mathletters}
\label{eq:metrics}
\begin{equation}
\label{eq:metric}
g_{\mu\nu} = \left[
\begin{array}{cccc}
\gamma & 0 & 0 & 0 \\
0 & -\alpha & 0 & 0 \\
0 & 0 & -\beta & f\sin\theta \\
0 & 0 & -f\sin\theta & -\beta\sin^2\theta
\end{array}
\right].
\end{equation}
Here $\alpha$, $\beta$, $\gamma$, and $f$ are functions of
$\xi^1=r$ only.
The inverse metric is
\begin{equation}
\label{eq:inverse_metric}
g^{\mu\nu} = \left[
\begin{array}{cccc}
1/\gamma & 0 & 0 & 0 \\
0 & -1/\alpha & 0 & 0 \\
0 & 0  & - \beta/(\beta^2+f^2) & f\csc\theta / (\beta^2+f^2) \\
0 & 0 & - f\csc\theta / (\beta^2+f^2) & -
\beta\csc^2\theta / (\beta^2+f^2) \\
\end{array}
\right] .
\end{equation}
Since $s_{\mu\nu}$ is a diagonal matrix,
it is a simple matter to find its inverse
\begin{equation}
\label{eq:inverse_of_symmetric_metric}
s^{\mu\nu} = \left[
\begin{array}{cccc}
1/\gamma & 0 & 0 & 0 \\
0 & -1/\alpha & 0 & 0 \\
0 & 0 & - 1 / \beta & 0 \\
0 & 0 & 0 & - 1 / \beta\sin^2\theta \\
\end{array}
\right] .
\end{equation}
\end{mathletters}This
satisfies
$s^{\mu\nu}s_{\mu\alpha} = \delta^\nu_\alpha$.
The function $\beta$ is assumed to be $\beta=r^2$.
In cases where $M \ll r$, the metric components
$\gamma$ and $\alpha$ take on the Schwarzschild form:
\begin{equation}
\label{eq:alpha_gamma}
\gamma = \frac{1}{\alpha} = 1 - \frac{2M}{r} ,
\end{equation}
so that $\alpha\gamma = 1$.
As we limit ourselves to a study of test-particle motion in weak-fields,
we will assume that it is always true that $\alpha$ and $\gamma$ take on
the Schwarzschild form.

Note that the only independent, non-zero component of
$a_{\mu\nu}$ is
$a_{23}= f\sin\theta$.
In the case of ``magnetic'' NGT, there is also a contribution from
$a_{10}= w$, where $w$ is a function of $r$ only.
It can be shown \cite{bib:Neil} that
when expanding about a spherically symmetric GR background,
the only solution for magnetic NGT that yields asymptotically flat
space is $w=0$.

{}From the definition of the NGT Christoffel symbols,
it is found that
\begin{mathletters}
\label{eq:Christoffels}
\begin{eqnarray}
\Christoffel{r}{(rr)}
&=& \frac{\alpha'}{2\alpha}
= -\frac{M}{r^2}\left(\frac{1}{1-2M/r}\right) \\
\Christoffel{r}{(\phi\phi)}
&=& \sin^2\theta\Christoffel{r}{(\theta\theta)}
= -\frac{r\sin^2\theta}{\alpha}
= -r\sin^2\theta\left(1-\frac{2M}{r}\right) \\
\Christoffel{r}{(tt)}
&=& \frac{\gamma'}{2\alpha}
= \frac{M}{r^2}\left(1-\frac{2M}{r}\right) \\
\Christoffel{\theta}{(r \theta)}
&=& \Christoffel{\phi}{(r \phi)}
= \frac{1}{r} \\
\Christoffel{\theta}{(\phi\phi)}
&=& -\sin\theta\cos\theta \\
\Christoffel{\phi}{(\theta\phi)}
&=& \frac{\cos\theta}{\sin\theta} \\
\Christoffel{t}{(rt)}
&=& \frac{\gamma'}{2\gamma}
= \frac{M}{r^2}\left(\frac{1}{1-2M/r}\right) .
\end{eqnarray}
\end{mathletters}A prime denotes differentiation with respect to $r$.
We have assumed that $M \ll r$, and hence limited the expressions
for the Christoffel symbols to their Schwarzschild form.

The determinant of the metric is given by
\begin{equation}
\label{eq:determinant}
g = \det(g_{\mu\nu})
= - (r^4+f^2)\sin^2\theta .
\end{equation}

In the static, spherically symmetric field, the skew field-strength
tensor $F_{[\mu\nu\lambda]}$ has only one independent,
non-zero component,
\[
F_{[\theta\phi r]} = \frac{1}{3}\partial_r a_{\theta\phi}
= \frac{1}{3}f'\sin\theta .
\]
On the other hand, ${{\bf g}^{[\lambda\nu]}}_{,\nu}$ vanishes:
\[
{{\bf g}^{[\theta\phi]}}_{,\phi}
= \frac{\partial}{\partial\phi}\left(
\frac{f}{\sqrt{r^4+f^2}}\right) \equiv 0 .
\]
{}From (\ref{eq:skew_force}), it follows that the skew force
$f_{[\alpha\sigma]}$ also has
one independent component:
\[
f_{[rt]} = \frac{d}{dr}\left(\frac{C_1\gamma f'}{\sqrt{r^4+f^2}}\right) .
\]
For convenience, we will set $C_1=1$ and $C_2=C_3=0$, as
these terms will not contribute to the motion for the case of static,
spherical symmetry.

In section~\ref{sec:Conservation_Laws}, it was found that
$J = r^2\dot\phi\sin\theta$ was a constant of the motion, corresponding
to the conservation of angular momentum per unit rest mass.
Using (\ref{eq:Christoffels}) and (\ref{eq:determinant}) in
(\ref{eq:equation_of_motion_massive}) yields the equations of motion for a
test-particle in the field of a static, spherically symmetric source:
\begin{mathletters}
\begin{eqnarray}
0 &=&
\frac{d^2r}{d\tau^2}
+ \frac{\alpha'}{2\alpha}\left(\frac{dr}{d\tau}\right)^2
- \frac{r\sin^2\theta}{\alpha}\left(\frac{d\phi}{d\tau}\right)^2
- \frac{r}{\alpha}\left(\frac{d\theta}{d\tau}\right)^2
+ \frac{\gamma'}{2\alpha}\left(\frac{dt}{d\tau}\right)^2
\nonumber \\
& & \mbox{}
\label{eq:r_component}
+ \frac{\lambda}{\alpha}\frac{dt}{d\tau}\frac{d}{dr}
\left(\frac{\gamma f'}{\sqrt{r^4+f^2}}\right) \\
0 &=&
\label{eq:theta_component}
\frac{d^2\theta}{d\tau^2}
+ \frac{2}{r}\frac{d\theta}{d\tau}\frac{dr}{d\tau}
- \sin\theta\cos\theta\left(\frac{d\phi}{d\tau}\right)^2 \\
0 &=&
\label{eq:t_component}
\frac{d^2t}{d\tau^2}
+ \frac{\gamma'}{\gamma}\frac{dt}{d\tau}\frac{dr}{d\tau}
+ \frac{\lambda}{\gamma}\frac{dr}{d\tau}\frac{d}{dr}
\left(\frac{\gamma f'}{\sqrt{r^4+f^2}}\right) .
\end{eqnarray}
\end{mathletters}The $\phi$ component is omitted, as it is already
integrated.

We can satisfy (\ref{eq:theta_component}) identically by
letting $\theta(\tau_0)=\pi /2$ and $\dot\theta(\tau_0) = 0$ for some
proper time $\tau_0$
(see \cite{bib:Papapetrou}, p.~71).
Orbits therefore lie in a plane and by choosing
$\theta(\tau_0) = \pi/2$, we have fixed that plane.
It follows that $J=r^2\dot\phi$.

We can write (\ref{eq:t_component}) as
\[
\frac{1}{\gamma}\frac{d}{d\tau}\left(\gamma\frac{dt}{d\tau}\right)
+ \frac{\lambda}{\gamma} \frac{d}{d\tau}\left(
\frac{\gamma f'}{\sqrt{r^4+f^2}}\right) = 0 .
\]
Since $\lambda$ is a constant, we conclude from this
that
\begin{equation}
\label{eq:dt_by_dtau}
E = \gamma\frac{dt}{d\tau}
+ \frac{\lambda \gamma f'}{\sqrt{r^4+f^2}}
\end{equation}
is a constant of the motion.
$E$ represents the energy at infinity per unit rest mass.
We see that, as was suggested earlier, although $\gamma\dot t$ is not a
constant of the motion, it does form one part of a conserved quantity.

Using these results, we can rewrite (\ref{eq:r_component}) as
\begin{eqnarray}
0 &=& \frac{d^2r}{d\tau^2}
+ \frac{\alpha'}{2\alpha}\left(\frac{dr}{d\tau}\right)^2
- \frac{J^2}{r^3\alpha}
+ \frac{\lambda}{\alpha\gamma}
\left(E-\frac{\lambda\gamma f'}{\sqrt{r^4+f^2}}\right)
\frac{d}{dr}\left(\frac{\gamma f'}{\sqrt{r^4+f^2}}\right) \nonumber \\
& & \mbox{}
\label{eq:r_motion_full}
+ \frac{\gamma'}{2\alpha \gamma^2}\left(E -
\frac{\lambda\gamma f'}{\sqrt{r^4+f^2}}\right)^2  .
\end{eqnarray}

In order to extract a useful result from this, we will make certain
simplifying assumptions.
There are two regimes to consider,
corresponding to $\mu r \gg 1$ (large~$r$) and
$\mu r \ll 1$ (small~$r$), respectively.
In both cases, we will assume $M \ll r$ and $M \ll 1/\mu$.

We treat first the case of large~$r$.
In this case, we are interested in the corrections to the
Newtonian gravitational force acting on the particle.
We therefore rewrite (\ref{eq:r_motion_full}) in terms of
$r(t)$:
\[
\frac{d^2r}{dt^2} + \frac{\alpha'}{2\alpha}\left(\frac{dr}{dt}\right)^2
- \frac{{J_N}^2}{r^3\alpha} + \frac{\gamma'}{2\alpha}
= - \frac{\lambda\gamma}{\alpha}
\left(E-\frac{\lambda\gamma f'}{\sqrt{r^4+f^2}}
\right)^{-1}\frac{d}{dr}\left(\frac{\gamma f'}{\sqrt{r^4+f^2}}\right) .
\]
Here, $mJ_N \equiv mr^2d\phi/dt$ is the Newtonian value of the
angular momentum.
We recognize the left-hand side as the usual GR contributions
to the equation of motion.

For large $r$, it can be shown \cite{bib:Neil} that
\[
f \approx \frac{sM^2}{3}\frac{e^{-\mu r}(1+\mu r)}{(\mu r)^{\mu M}} ,
\]
where $s$ is an arbitrary
constant and $M$ is the Schwarzschild mass of the source.
The constant $1/\mu$ is a fundamental constant in the NGT, and
represents the range of the skew
components $a_{\mu\nu}$.
Assuming $\mu M \ll 1$, we can write $(\mu r)^{\mu M} \sim 1$,
leaving
\begin{equation}
\label{eq:f}
f \approx \frac{sM^2}{3} e^{-\mu r} (1 + \mu r) .
\end{equation}

To our order of approximation,
\[
\frac{\lambda\gamma}{\alpha}
\left(E-\frac{\lambda\gamma f'}{\sqrt{r^4+f^2}}
\right)^{-1}\frac{d}{dr}\left(\frac{\gamma f'}{\sqrt{r^4+f^2}}\right)
\approx E\lambda
\frac{d}{dr}\left(\frac{\gamma f'}{\sqrt{r^4+f^2}}
\right)
\approx \frac{E\lambda
s M^2 \mu^2}{3} \frac{e^{-\mu r}(1+\mu r)}{r^2} .
\]
Therefore, the radial equation of motion may be written
\begin{equation}
\label{eq:look_at_this}
\frac{d^2r}{dt^2} - \frac{{J_N}^2}{r^3} = - \frac{M}{r^2}
- \frac{E\lambda
sM^2\mu^2}{3}\frac{e^{-\mu r}(1+\mu r)}{r^2} ,
\end{equation}
where we have assumed that the particle is moving slowly, so
that $dr/dt \ll 1$.
For a spin $1^+$ particle exchange, as is obtained in the linear
approximation of the NGT \cite{bib:Moffat_NGT_2},
$\lambda < 0$,
yielding a repulsive Yukawa force in (\ref{eq:look_at_this}).

At the other end of the spectrum, we have the case where $r$ can be
taken as small.
More precisely, this is the region where $\mu r$ and $M/r$ are both
small.
We are more interested here in the orbit of the particle.
It is interesting to take a different approach:
taking $\theta = \pi/2$ and $\dot\theta = 0$, the
normalization condition $s_{\mu\nu}\dot\xi^\mu\dot\xi^\nu=1$
can be written
\[
1 = \gamma \left(\frac{dt}{d\tau}\right)^2
- \alpha\left(\frac{dr}{d\tau}\right)^2
- r^2\left(\frac{d\phi}{d\tau}\right)^2
= \gamma\left(\frac{dt}{d\tau}\right)^2
- \alpha J^2 \left[\left(\frac{du}{d\phi}
\right)^2 + \frac{1}{r^2\alpha}\right] ,
\]
where $J = r^2 d\phi/d\tau$, $u=1/r$, and where we have written
$r=r(\phi)$.
Using (\ref{eq:dt_by_dtau}), this becomes
\[
\left(\frac{du}{d\phi}\right)^2
= \frac{1}{\alpha\gamma J^2}
\left(E - \frac{\lambda\gamma f'}{\sqrt{r^4+f^2}}\right)^2
- \frac{1}{r^2\alpha} - \frac{1}{\alpha J^2} .
\]
Differentiating this with respect to $\phi$ gives the equation
describing the orbit of the particle:
\begin{equation}
\label{eq:orbit_equation}
\frac{d^2u}{d\phi^2}
= - \frac{r^2}{2}\frac{d}{dr}\left[\frac{1}{\alpha\gamma J^2}
\left(E - \frac{\lambda\gamma f'}{\sqrt{r^4+f^2}}\right)^2
- \frac{1}{\alpha J^2} - \frac{1}{r^2\alpha}\right] .
\end{equation}

Thus far, the treatment has been entirely general.
We can specialize to the
case where~$\mu r$ and $M/r$ are small.
It can then be shown \cite{bib:Neil} that
\[
f \approx \frac{sM^2}{3}\left(1 - \frac{1}{2}(\mu r)^2
+ \frac{2M}{r}\right) .
\]
Typically, the term $2M/r$ in the parenthesis will dominate the
term $(\mu r)^2$.
However, for interest sake we will keep both terms.
Taking the Schwarzschild forms $\gamma = 1/\alpha = 1-2M/r$, we
arrive at the orbit equation:
\begin{equation}
\label{eq:lowest_order_orbit}
\frac{d^2u}{d\phi^2} + u
= \frac{M}{J^2}\left(1 + \frac{E\lambda sM\mu^2}{3}\right)
 + 3Mu^2 + \frac{8\lambda sM^3u^3}{3J^2} .
\end{equation}
This equation is similar to the orbit equation from
GR (see \cite{bib:Weinberg}, p.~186),
the only differences being
the last term on the right-hand side and
the factor multiplying the first term on the right-hand side.
The former term represents the lowest-order NGT correction to the GR
result.

\section{Conclusions}

In \cite{bib:Wheeler}, it is argued that an analysis of particle motion
in the strong-field regime from the point of view of the matter
response equation (see \cite{bib:Papapetrou_2}) is inconclusive due to
the uncertain nature of the stress tensor, $T^{\mu\nu}$.
A similar judgement is passed on an analysis of the type performed in
\cite{bib:EIH}, since ``If singularities are tolerated in general
relativity, then the completeness of this theory is even more thoroughly
shattered.''
As NGT is claimed to be a candidate for a non-singular theory
of gravitation, an analysis of particle motion by means of point-like
singularities is even more questionable.
However, these types of analyses can be performed in the weak-field
regime.

Far be it for us to discuss the validity of such criticism,
we have preferred to avoid the problem of solving the motion
of particles in strong gravitational fields
by restraining ourselves to
an analysis of test-particle motion strictly in the weak-field regions
of the static, spherically-symmetric solution of the field equations.
The particular choice of the Lagrangian $L_{\rm GR}$ leading to the
familiar left-hand sides of (\ref{eq:equation_of_motion_massive})
and (\ref{eq:equation_of_motion_massless})
is guided by the matter response equation, which in NGT
may be written (see \cite{bib:Legare_Moffat} for a derivation of the
matter response equation):
\begin{equation}
\label{eq:matter_response}
{{\bf T}^{(\beta\rho)}}_{,\rho}
+ s^{\beta\lambda}a_{\mu\lambda}{{\bf T}^{[\mu\rho]}}_{,\rho}
+ \Christoffel{\beta}{\mu\nu}{\bf T}^{\mu\nu} = 0 ,
\end{equation}
where
\[
\Christoffel{\beta}{\mu\nu}
= \frac{1}{2}s^{\beta\alpha}
\left(g_{\alpha\nu,\mu} + g_{\mu\alpha,\nu}
- g_{\mu\nu,\alpha}\right) .
\]
The resemblance of this object to the Christoffel symbol defined in
(\ref{eq:Christoffel_symbol}) is obvious.
The matter response equation
is a result of the full NGT field equations,
valid in both strong- and weak-field regimes.
Should the ``correct'' stress tensor $T^{\mu\nu}$ be known, it is
reasonable to conclude that the predicted test-particle equation of
motion would contain the Christoffel connection.
For instance, we might expect that in the case of massless particles
this would lead to (\ref{eq:equation_of_motion_massless}).

However, things are not so simple: according to Wheeler \cite{bib:Wheeler},
the arguments of the previous paragraph are conclusive only in the
weak-field regions of spacetime.
Taking the weak-field limit to be the deciding factor, we find the number
of candidates for the title of ``equation of test-particle motion''
multiplies quickly.
For instance, the left-hand sides of (\ref{eq:equation_of_motion_massive})
and (\ref{eq:equation_of_motion_massless})
can be written $\nabla_{\dot\xi}[\xi]^\beta$, where $\nabla_\mu$ is the
connection defined with respect to the Christoffel symbols.
Although we cannot obtain the resulting equations of motion from an
action principle, we could have just as easily postulated that the
aforementioned connection be defined with respect to some other set of
connection coefficients.
There are two obvious possibilities in the NGT: $\Gamma^\beta_{\mu\nu}$
and $W^\beta_{\mu\nu}$.
The latter is unconstrained, while the former is constrained to have
a vanishing torsion vector: $\Gamma^\beta_{[\mu\beta]} = 0$.
In the weak-field limit, both of these connections reduce to the
Christoffel symbols, and therefore lead to the same equation of motion.
These possibilities will be the subject of a future study.

Keeping the previous arguments in mind, we have brought forth three
possible candidates to couple the test-particle quantities to the
skew field of NGT.
It is found that, in this scheme, massless particles would not directly
couple to the skew field.

Of the three proposed
candidates, only one is found to generate any interaction
at all in a static, spherically-symmetric field.
The correction to the Newtonian gravitational force acting on the
particle is worked out in the weak-field regime and is described by
a repulsive Yukawa force.
The correction to the orbit of a particle is also calculated; the lowest-order
NGT correction is a $1/r^3$ term.

\acknowledgements

We would like to thank M. Clayton, N.J. Cornish, and
T. Demopoulos
for their help and for many stimulating
discussions.
The form of the coupling (\ref{eq:dual_coupling}) was
suggested by M. Clayton.
This work was supported by the Natural Sciences and Engineering
Research Council of Canada.
We would like to thank the University of West Indies, Cave Hill
Campus, for their hospitality during the course of this work.
J. L\'egar\'e would like to thank the Government of Ontario for their
support of this work.

\end{document}